\documentclass[aps,prl,twocolumn,amsmath,amssymb,floatfix]{revtex4}

\usepackage{amsmath}
\usepackage{graphicx}
\usepackage{epsfig}
\usepackage[utf8]{inputenc}
\usepackage{bm}
\usepackage{hyperref}

\def\H1{\widehat{H}_1}

\newcommand{\pd}{\partial}
\newcommand\vvdots{\vphantom{\int^0}\smash[t]{\vdots}}

\begin{document}

\title[]{Quantum choppers}

\author{Mikhail Pletyukhov$^{1}$, Kim G. L. Pedersen$^{1}$, Vladimir Gritsev$^{2}$ }

\affiliation{
$^1$Institute for Theory of Statistical Physics and JARA -- Fundamentals of Future Information Technology, 
RWTH Aachen University, 52056 Aachen, Germany \\
$^2$Institute for Theoretical Physics, Universiteit van Amsterdam, Science Park 904,
Postbus 94485, 1098 XH Amsterdam, The Netherlands
}
\begin{abstract}
An optical chopper periodically interrupts a classical light beam. We propose a realizable quantum version of the optical chopper, where the time-periodic driving of the light-matter coupling is achieved through a nonlinear three wave mixing element. We theoretically investigate how our scheme can be used for the controllable shaping of few photon light. Using Floquet dynamics, we find strong periodic modulations of the transmission and reflection envelopes in the scattered few-photon pulses, including photon compression and blockade, as well as dramatic changes in the quantum light statistics. Our theoretical analysis allows us to explain these non-trivial phenomena as arising from non-adiabatic memory effects. 
\end{abstract}

\maketitle

\emph{Introduction.---}
Optical choppers and shutters~\cite{Choppers} are ubiquitous optical instruments, famous, perhaps, for their application in the first non-astronomical speed of light measurements by Hippolyte Fizeau in 1849~\cite{Fizeau}, and used today for e.g. speed or rotation measurements, light exposure control, and off-frequency noise filtering. The prototypical chopper uses a rotating wheel with holes that periodically block the incident light beam, with the hole radius to beam width ratio controlling the waveform of the chopped light~\cite{Choppers2}.

Recent advances in nanophotonic technologies allow for a controllable manipulation of light-matter couplings~\cite{Nori}, and we show in this Letter, how a time-periodic coupling can be designed and used to implement an optical chopper down to the quantum single-photon level -- a quantum chopper. The future applications of quantum choppers might be as diverse as those of its classical counterpart, especially, due to the increasing interest in periodiccaly driven open or closed quantum systems. Theoretically, various effective Floquet models~\cite{GH} already describe photon-assisted tunneling in quantum wells and the dynamics of quantum open systems. Experimentally, electron-spin and nuclear magnetic-spin resonances access the time-evolution of spinful particles placed time-dependent oscillating magnetic fields.
In quantum information dynamical decoupling schemes~\cite{VL,Ban,VKL,Z,KLV} and their refinements~\cite{KL,U}, periodic sequences of fast and strong symmetrizing pulses are used to reduce some parts of a system-bath interaction, which cause decoherence. The Floquet systems also naturally appear in digital quantum computation schemes \cite{F15}, and in recent years periodic perturbations have been used as a flexible experimental tool to engineer new, synthetic  phases of matter not accessible in equilibrium systems \cite{F4,F5,F6,F7,F16,F8,F9,F10,F11,F12,F13,DP}.

\begin{figure}[ht]
    \includegraphics[width=\columnwidth,angle=0]{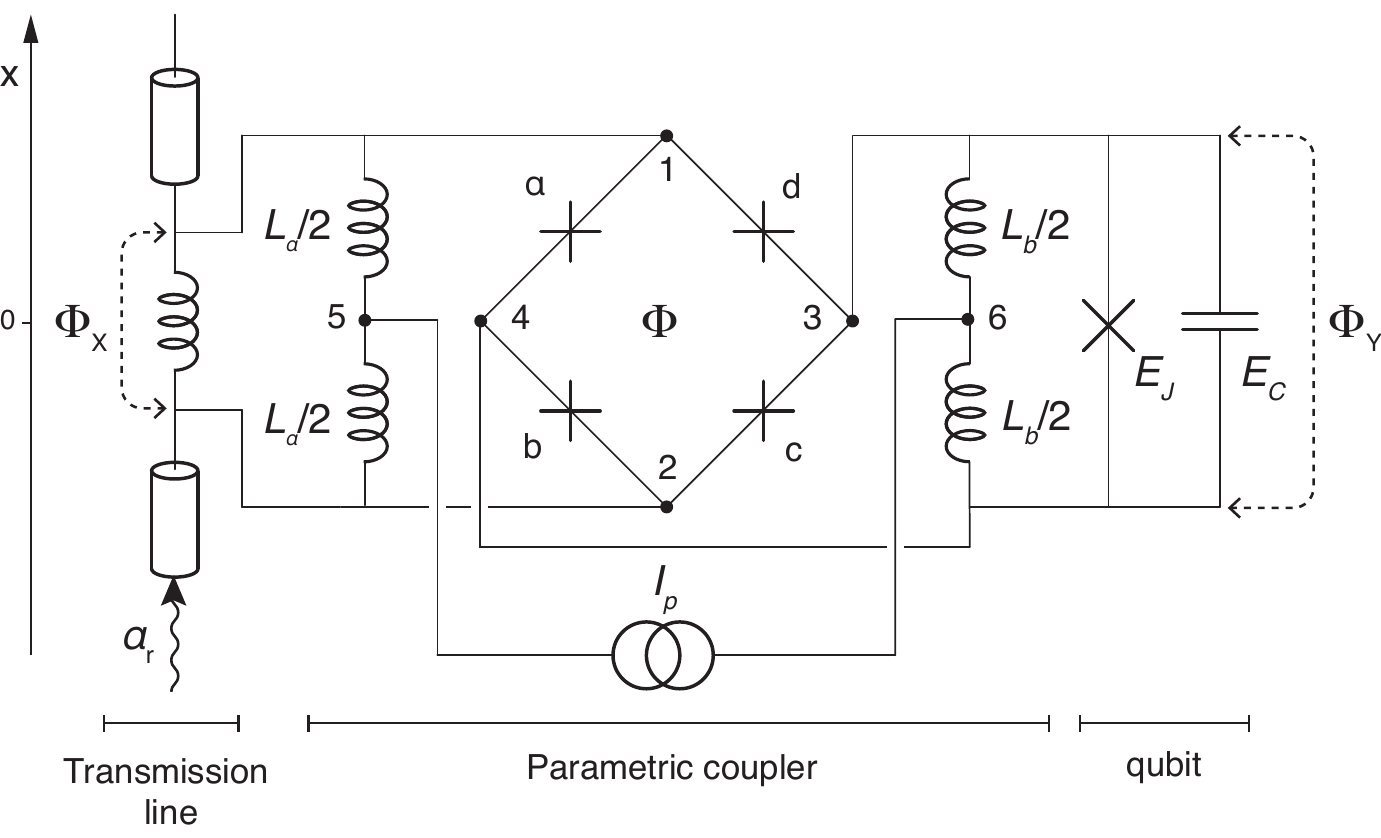}
    \caption{A microwave variant of the quantum chopper scheme. The system consists of the transmon qubit (right part) coupled to the transmission line (left part) via a nonlinear element --- the Josephson ring modulator (central part). The coupling between the qubit and the transmission line is modulated by the time-dependent current $I_p (t)$.  } \label{scheme}
\end{figure}

Our quantum chopper scheme, shown in Fig.~\ref{scheme}, uses the Josephson ring modulator (JRM) \cite{Devoret} which is based on the Josephson junction Wheatstone bridge. While the JRM has been originally conceived as a parametric converter and afterwards employed \cite{Hafchem} as a tool to generate a chemical potential for photons, we suggest to use it as a nonlinear element that controls a time-dependent coupling between a transmission line and a qubit. The role of the JRM is to produce an analogue of the parametric down conversion, such that one of the modes becomes classical and can be driven periodically by an external classical current. We describe the system in Fig.~\ref{scheme} in terms of scattering of microwave photons, which propagate in the transmission line, from the qubit.

We develop a general multi-photon Floquet scattering formalism for time-periodic light-matter couplings and apply it to describe the quantum properties of light, like transmittance, reflectance, first- and second-order coherences. This allows us to demonstrate the quantum chopper as a tool for pulse shaping of individual photons.

We find that for weakly coherent incident pulses a time-periodic modulation of the coupling between photons and the cavity causes considerable modifications to envelopes of the scattered photons. In some driving protocols, the field can be even locally quenched, with the envelope featuring nodes, which is a quantum analogue of the classical light chopping. We also find a significant effect of the modulation on the photon's statistics. Thus, for reflected photons on resonance we observe alternating patterns of bunching and antibunching which periodically change in time --  this contrasts with the purely antibunching behavior in the case of constant coupling.

\emph{Setup.---}  
We propose a scheme for realizing quantum optical chopping (shown in Fig.~\ref{scheme}) consisting of a transmission line with linear dispersion of propagating microwave photons, a parametric coupler, and a transmon qubit. Following \cite{Devoret} we introduce the node flux variable $\Phi_{1,2,3,4}$ in order to describe the parametric coupler using the JRM pierced by the external flux $\Phi$. Its energy consists of Josephson energies  $-E'_J \cos (\delta_i +\frac{\pi \Phi}{2 \Phi_0})$ stored in the four Josephson junctions ($i=a,b,c,d$) forming the ring, where $\delta_i$ are the phase differences across the corresponding junctions [$\delta_a = \frac{2 \pi}{\Phi_0} (\Phi_4 - \Phi_1)$, etc.] and $\Phi_0 = \frac{h}{2e}$ is the flux quantum; 
of inductive energies $\frac{1}{L_a} (\Phi_5 - \Phi_1)^2+ \frac{1}{L_a} (\Phi_5 - \Phi_2)^2$ and $\frac{1}{L_b} (\Phi_6 - \Phi_3)^2+ \frac{1}{L_b} (\Phi_6 - \Phi_4)^2$ (using the additional node flux variables $\Phi_{5,6}$); and of coupling energy $I_p (\Phi_5 - \Phi_6)$ to the AC current source $I_p (t)$.  Introducing modes $\Phi_X = \Phi_2 - \Phi_1$, $\Phi_Y = \Phi_3 - \Phi_4$, $\Phi_Z = \Phi_3 + \Phi_4 - \Phi_1 - \Phi_2$,  $\Phi_l = \Phi_5 - \frac{\Phi_1 +\Phi_2}{2}$, $\Phi_r = \Phi_6 - \frac{\Phi_3 +\Phi_4}{2}$, one achieves \cite{Devoret,seeSupple} the nonlinear three-mode mixing $\Phi_X$, $\Phi_Y$, and $\Phi_Z$, while the modes $\Phi_l$ and $\Phi_r$ decouples from the rest. Expanding the nonlinear coupling in $\Phi_{X,Y,Z}/\Phi_0 \ll 1$ up to cubic order and choosing the optimal point $\Phi = \Phi_0/2$ to maximize the coupling constant of the three-mode mixing we obtain
\begin{equation}
H_{JRM} = \lambda \Phi_{X}\Phi_{Y}\Phi_{Z} + \sum_{s=X,Y,Z} \mu_s \Phi_{s}^{2} - \frac12 I_p \Phi_Z,
\label{JRM}
\end{equation}
where ${\lambda=-2\sqrt{2}\pi^{3}E'_{J}/\Phi_{0}^{3}}$, ${\mu_Z=\sqrt{2}\pi^{2}E'_{J}/\Phi_0^{2}}$, and ${\mu_X = \mu_Z + 1/(2 L_a)}$, ${\mu_Y = \mu_Z + 1/(2 L_b)}$. 

The mode $\Phi_{Z}$ is coupled to the AC current source $I_p (t)$ with frequency $\Omega$ and a sufficiently large amplitude. Therefore, $\Phi_Z$ becomes classical and fixed to the value $\Phi_Z (t) = I_p (t)/(4 \mu_Z)$. The mode $\Phi_{Y}$ coincides with the flux variable of the qubit, which is characterized by the Josephson ($E_J$) and capacitive ($E_c$) energies. In the transmon limit $E_J \gg E_C$, the qubit is represented by a nonlinear single-mode cavity in terms of the cavity operators $b$ and $b^{\dagger}$, with a cavity frequency $\omega_c = \frac{1}{\hbar} \sqrt{8 E_J E_C}$ and a Kerr nonlinearity $\frac{U}{2} b^{\dagger \, 2} b^2 = - \tfrac{1}{2} E_C b^{\dagger \, 2} b^2$. 

The mode $\Phi_X$ coincides with the flux variable of the transmission line in its middle. Quantizing the transmission line, we express ${\Phi_X = \int d k \frac{f_k}{\sqrt{2}} (a_{r,k} + a_{l,k})+h.c.}$ in terms of right ($a_{r,k}$) and left ($a_{l,k}$) microwave photons propagating along the $x$-axis with a constant phase velocity, $v$. The coefficients $f_k$ depend on the parameters of the transmission line and on the photonic frequencies $\omega = v |k|$. Assuming that the frequency $\omega_0$ of the injected (right-moving) photons in the mode $k_0$ is commensurate with the qubit's transition frequency, $\omega_c$, and that $\omega_0$ is much larger than the drive frequency $\Omega$, the qubit's decay rate $\Gamma$, and the detuning $\delta = \omega_0 - \omega_c$, we neglect the frequency dependence of $f_k$, setting $f_k  \approx f_{k_0}$, and perform the rotating wave approximation in the product ${\Phi_X \Phi_Y = \Phi_X \sqrt{ 2 E_c / E_J} (b+b^{\dagger})}$. Thus, we obtain a time-dependent coupling $V (t) = g (t) \int d k (a_k^{\dagger} b + b^{\dagger} a_k)$  between the transmission line and the qubit, where $g (t) = \lambda f_{k_0} \sqrt{2 E_c / E_J} I_p (t) /(4 \mu_Z)$, and $a_k = \frac{1}{\sqrt{2}} (a_{r,k}+a_{l,-k})$ is the even combination of the right and left propagating photonic fields. Complementing $V (t)$ by the free Hamiltonian
${H_0 = \int d k \hbar v k (a_k^{\dagger} a_k + \tilde{a}_k^{\dagger} \tilde{a}_k) + \hbar \omega_c b^{\dagger} b}$, where $\tilde{a}_k =  \frac{1}{\sqrt{2}} (a_{r,k}-a_{l,-k})$ is the odd fields' combination decoupled from the qubit, we obtain the Hamiltonian $H (t) = H_0 +V (t)$ of our model. In the following we use units where $\hbar=v=1$.

\emph{Floquet scattering formalism.---} An extension of the scattering formalism for time-periodic Hamiltonians was originally proposed in Ref.~\cite{PM} for the calculation of above-threshold-ionization spectra. Remarkably, it offered an effectively time-independent description of the quasistationay limit in terms of the  Floquet states. Later, similar scattering approaches have been developed for single-particle scattering~\cite{LR,ER} and many body scattering of non-interacting particles~\cite{Moskal-b,Moskal1}. One recent study~\cite{BC} dwells on the calculation of inelastic scattering rates in driven interacting systems. 

Here we elaborate on the Floquet scattering formalism, particularly adapting it to problems of multi-particle scattering of microwave photons in transmission lines interacting with nonlinear cavities. We present a systematic way of computing $N$-photon scattering operators $S_N$ for a time-periodic interaction $V (t) = \sum_m V^{(m)} e^{-i m \Omega t}$ in the Floquet-extended Hilbert space, thereby generalizing the approach of Ref.~\cite{PG} for time-independent couplings. 

Let us briefly review the basic ideas of scattering theory. Suppose that at time $t_0 \to - \infty$ we inject $N$ photons into the transmission line, while the cavity is empty. In the second quantization, this incoming state is given by ${|p \rangle \equiv | \{ k_j \} , l=0 \rangle = \left( \prod_{j=1}^N a_{k_j}^{\dagger} \right) |\mathrm{vac}, l=0 \rangle}$, where the vacuum state $|\mathrm{vac} \rangle $ of the transmission line is defined by $a_k | \mathrm{vac} \rangle = \tilde{a}_k |\mathrm{vac} \rangle = 0$, and $|l \rangle$ is the photon number state of the cavity, $b^{\dagger} b | l \rangle = l | l \rangle$. The energy of the incoming state equals $\varepsilon_p = \sum_{j=1}^N k_j$, and in case of the time-independent interaction $V$ this energy is conserved in the following sense: matrix elements $S_{N;p'p} = \langle p' | S_N | p \rangle$ of the scattering operator $S_N$, which defines an outgoing scattering state  $S_N | p \rangle$ at $t \to \infty$, appear to be proportional to delta functions $\delta (\varepsilon_{p'} - \varepsilon_{p})$. Moreover, $S_{N;p'p} = \delta_{p'p} - 2 \pi i \delta (\varepsilon_{p'} - \varepsilon_{p}) T_{N; p'p} (\varepsilon_{p})$, where $T_{N;p'p} (E)$ is the energy-dependent $T$ matrix containing all the information about scattering off the cavity. A systematic way of computing $T_N (E)$ has been developed in \cite{PG} for scatterers with an arbitrary level structure and transition matrix elements.

For a time-periodic interaction, $V (t)$, elements of the $S$ matrix are modified~\cite{seeSupple} to ${S_{N;p'p} = \delta_{p'p} - 2 \pi i \sum_{m'} \delta (\varepsilon_{p'} -m' \Omega - \varepsilon_{p}) T_{N; p'p}^{(m')} (\varepsilon_{p})}$, where now the energy of an incoming state is conserved modulo an integer number of the drive frequency quanta. The $N$-photon $T$ matrix bears an additional integer-valued Floquet index $m'$, and its elements are found from the equation 
\begin{equation}
T_{p'p}^{m'm} (E) = V_{p'p}^{m'm} + V_{p' q'}^{m' n'}
\left[ \frac{1}{E - H' +i 0^+} \right]_{q'q}^{n' n} V_{qp}^{nm}
\label{TVV}
\end{equation}
for the operator $T = \sum_{N=1}^{\infty} T_N$, which is defined in the Floquet-extended Hilbert space, by setting $m=0$. In Eq.~\eqref{TVV}, we implicitly assume summations (integrations) over repeated discrete (continuous) indices; $H' = H - i \pd_\tau$ is the Floquet Hamiltonian with the operator $i \pd_{\tau}$ defined by $i \pd_{\tau} |m \rangle = m \Omega |m \rangle$ in terms of the Floquet states $|m \rangle = e^{-i m \Omega \tau}$, such that $\langle m' | m \rangle =\int_0^{T} \frac{d \tau}{T} e^{i (m'-m) \tau}= \delta_{m'm}$, and the Floquet-Hilbert space is spanned by $|p \rangle \otimes |m \rangle$.

Realizing that \eqref{TVV} has the form conventional in scattering theory, ${T (E) = V (E - H' +i 0^+)^{-1} V}$, we repeat the arguments of Ref.~\cite{PG} and obtain (almost) identical diagrammatic rules allowing us to calculate all $T_N$ explicitly~\cite{seeSupple} (their only modification consists in additional summations over Floquet indices). This, in turn, leads us to a closed form expression for $S_N$ conveniently written in the time representation
\begin{align}
S_N &= \hat{1}_N + (-i)^{2N} \int d t_1 \ldots d t_{2N} \Theta (t_1> \ldots > t_{2N}) \nonumber \\
&\times e^{i (\varepsilon_{p'} -\varepsilon_p) t_1}  P_0 \left( \vvdots \, V (t_1) e^{-F (t_1)} e^{F (t_2)} V (t_2) e^{-F (t_2)} \right. \nonumber \\ 
& \times \left. \ldots V (t_{2N-1}) e^{- F (t_{2 N-1})} e^{F (t_{2 N})} V (t_{2N}) \, \vvdots \right) P_0.
\label{SNmain}
\end{align}
Here ${P_0 = |l=0 \rangle \langle l=0|}$ is a projector onto the dark state of the cavity, $H'_0 = H_0 - i \pd_{\tau}$ is the free Floquet Hamiltonian, and  the operator $F (t) = i (H_0 +  \Sigma^{(0)} - \varepsilon_p) t + F_{\mathrm{osc}} (t)$ is given in terms of the time-dependent cavity's self-energy ${\Sigma (t) = - i   b^{\dagger} b \Gamma (t) = \sum_m \Sigma^{(m)} e^{- i m \Omega t}}$, ${\Gamma (t) = \pi g^2 (t)}$, and ${F_{\mathrm{osc}} (t) = - \sum_{m \neq 0} \Sigma^{(m)} e^{- i m \Omega t}/(m \Omega)}$. The symbol $\vvdots (\ldots ) \vvdots$ denotes a modified normal ordering, which, as compared to the conventional normal ordering, also shifts $H_0$ by virtual energies arising from the commutations of field operators contained in $V$ with $H_0$.

\emph{Results.---}Eq.~\eqref{SNmain} represents the central theoretical result of this Letter.
Assuming a weakly coherent initial signal in the right-moving mode $a_{r,k_0}$, on the basis of \eqref{SNmain} we establish first and second order coherences $g_{rr}^{(1,2)}$ and $g_{ll}^{(1,2)}$ corresponding to the scattered right (or transmitted) and left (or reflected) fields, respectively. 

The first-order coherences amount to ${g_{rr}^{(1)} (\tau_c) = |1 + A (\tau_c)|^2}$ and ${g_{ll}^{(1)} (\tau_c) = |A (\tau_c)|^2}$, where ${A (\tau_c) = -  w (\tau_c) W (\tau_c)}$ is a periodic function of the reduced central time ${\tau_c = (t - |x|)\! \mod T \in [-T/2 , T/2]}$. It is expressed via ${w (\tau_c) =  \sqrt{\pi} g (\tau_c) e^{-f_1 (\tau_c)}}$ and ${W (\tau_c) = \int_{-\infty}^{\tau_c} d t' e^{f_1 (t')} \sqrt{\pi} g (t')}$, where ${f_1 (t) = \langle l=1 | F (t) |l=1 \rangle}$. The functions $A (\tau_c) \equiv r (\tau_c)$ and $1+A(\tau_c) \equiv t (\tau_c)$ have the physical meaning of envelopes for the reflected and transmitted field. For a time independent interaction they reduce to the familiar reflection $r = -\frac{i \Gamma}{\delta + i \Gamma}$ and transmission $t = \frac{\delta}{\delta +i \Gamma}$ coefficients. The amplitude modulations $r (\tau_c)$ and $t (\tau_c)$ obey the following normalization condition $\int_{-T/2}^{T/2} \frac{d \tau_c}{T} (|r (\tau_c)|^2 + |t (\tau_c)|^2) =1$, which is equivalent to the unitarity of a single photon scattering matrix in the Floquet-Hilbert space.

\begin{figure}[tb]
	\centering
	\includegraphics[width=\columnwidth]{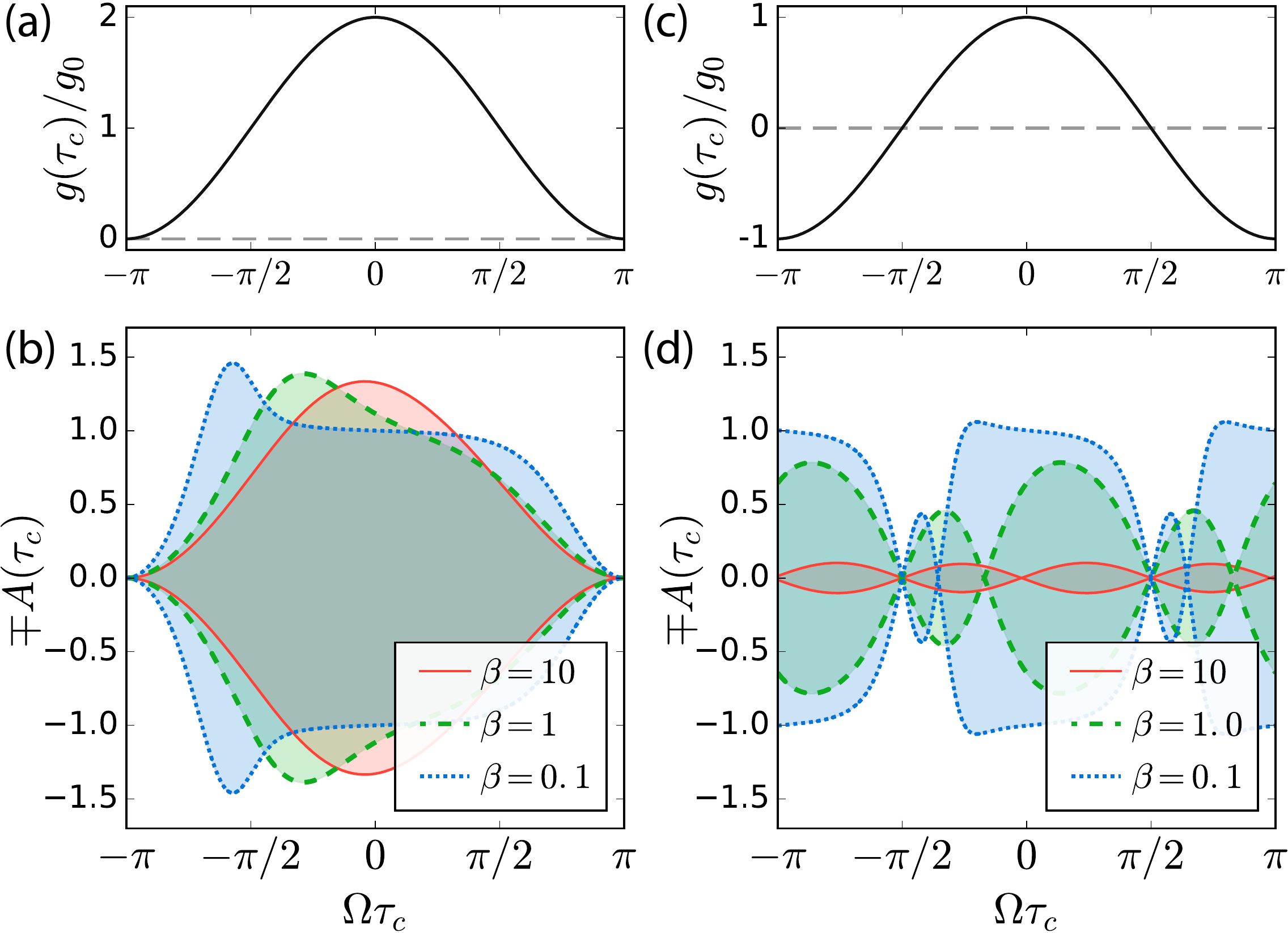}
	\caption{(Color online) The envelopes of the reflected field. \textbf{(a)} The ``on-off'' cosine signal, and the resulting \textbf{(b)} envelope as a function of the central time $\tau_c$ for various driving speeds. Note the perfect transmission ($A=0$) when the coupling is quenched. \textbf{(c)} The ``sign change'' cosine signal and the resulting \textbf{(d)} envelope. The envelope repeats itself after a half period, and in addition to the two coupling quench nodes at $\Omega \tau_c = \pm \pi/2$ an extra node develops at $\Omega \tau_c \approx -\pi/2$ (at slow drive) and moves towards $\tau_c = 0$ (at fast drive).}
	\label{fig:g1}
\end{figure}

We apply these general results to two coupling modulation protocols:
1) ``on-off'' $g (t) = g_0 (1+\cos \Omega t)$; and 2) ``sign change'' $g (t) = g_0 \cos \Omega t$. In the ``on-off'' protocol the coupling strength is periodically quenched to zero [Fig.~\ref{fig:g1}(a)], while in the ``sign change'' protocol, the sign of $g (t)$ changes after crossing zero [Fig.~\ref{fig:g1}(c)]. The major difference between the two protocols is that the former yields a $2 \pi$-periodic modulation of a field's amplitude [Fig.~\ref{fig:g1}(b)], while the latter yields a $\pi$-periodic one [Fig.~\ref{fig:g1}(d)]. 

For a time independent interaction, a single photon on resonance ($\delta=0$) is fully reflected ($r=-1$), regardless the value of the coupling strength. Should the adiabaticity condition $|\dot{g} (t)/ g(t)| \ll \sqrt{\delta^2 +\Gamma^2 (t)}$ be fulfilled at every time $t$ for a time periodic interaction, we would expect the reflection amplitude $r (t)$ to follow $\Gamma (t)$ instantaneously, also showing (almost) full reflection in the resonant case (the small fraction $|\dot{g} (t)/(g (t) \Gamma (t))|^2$ of the transmitted photon's probability density can be obtained from the first adiabatic correction). However, the adiabaticity condition is strongly violated for the two protocols. Moreover, at certain time instants the coupling strength in both of them is quenched, implying a momentary decoupling of microwave photons from the cavity and hence full transmission at these time instants. Since we are dealing with an open quantum system, this qualitative picture becomes even more complicated due to memory effects, and the non-adiabatic behavior can be explained as a sum over histories. Each history has the photon entering the cavity at some initial time, $\tau_i$, and leaving at some later time, $\tau_f$, with an amplitude $g(\tau_i) g(\tau_f)$, and a weight determined by the decay probability of the photonic state in the cavity, $\exp( - \int_{\tau_i}^{\tau_f} \Gamma(\tau) \mathrm{d}\tau)$. The reflection coefficient at $\tau_f$, given by the sum over initial times $\tau_i$, is highly influenced by the evolution within a memory window set by the decay rate of the cavity. 

In the ``on-off'' protocol the memory window is largest for final times after the $\Omega \tau_c = -\pi$ node, meaning that the photon remains longer in the cavity and is released shortly after when the coupling strength is sufficiently increased, producing a spike in the reflection coefficient that overshoots unity [Fig.~\ref{fig:g1}(b)].
In the ``sign change'' protocol memory effects create an additional node, that is absent in $g (t)$, close to $\Omega \tau_c = -\pi/2$ for slow drive and moving towards ${\tau_c = 0}$ for faster drives [Fig.~\ref{fig:g1}(d)]. For times shortly after the $-\pi/2$ node of $g (t)$ the memory window includes histories with amplitudes of opposite signs, and their competition creates this additional node. All this shows how different protocols not only chop the wavepacket of the incoming photon, but also significantly alter its form.

The resulting envelopes strongly depend on the normalized frequency $\beta = \Omega / \Gamma^{(0)}$, where $\Gamma^{(0)}$ is the zeroth harmonic of $\Gamma(t)$.
For the fast drive $\beta \gg 1$, we obtain ${A (\tau_c) \approx - \frac23 (1+\cos \Omega \tau_c)}$ in the ``on-off'' protocol, which means that the reflected pulse follows $g (\tau_c)$, not $\Gamma (\tau_c)$;  and ${A (\tau_c) \approx - \frac{1}{\beta} \sin 2 \Omega \tau_c}$ following $\Gamma^{(0)} \tau_c - f_1 (\tau_c)$ in the ``sign change'' protocol. In the second case, $A (\tau_c)$ is negligibly small, so that we have (almost) full transmission despite the resonance --- this effect is in sharp contrast to its non-driven counterpart. For the slow drive $\beta \ll 1$, the adiabaticity condition is fulfilled at least within some range around $\tau_c =0$, and this accounts for the formation of the plateau $A (\tau_c) \approx - 1$, resembling full reflection in the non-driven resonant case.

The second order coherences manifest nonlinear effects quantified by the value of $U$. They depend on the central time $\tau_c$ and the delay time $\tau_d>0$, being periodic in $\tau_c$ and aperiodic in $\tau_d$. In the explicit form, ${g_{ll}^{(2)} (\tau_c +\tau_d , \tau_c) = |1+ B (\tau_c , \tau_d)/(r(\tau_c+\tau_d) r (\tau_c)) |^2}$, with
$B (\tau_c , \tau_d) = -i  U w (\tau_c +\tau_d) w (\tau_c)  \int_{-\infty}^{\tau_c} d t' e^{i U (t' - \tau_c)} W^2 (t'). $ 
For time independent interaction this reduces to ${B (\tau_d) = r^2 U/(2 \delta+ 2 i \Gamma -U) e^{(i \delta - \Gamma) \tau_d}}$. The function $g_{rr}^{(2)}$ is obtained from $g_{ll}^{(2)}$ by replacing $r (\tau_c)$ with $t (\tau_c)$.

\begin{figure}[tb]
	\centering
	\includegraphics[width=\columnwidth]{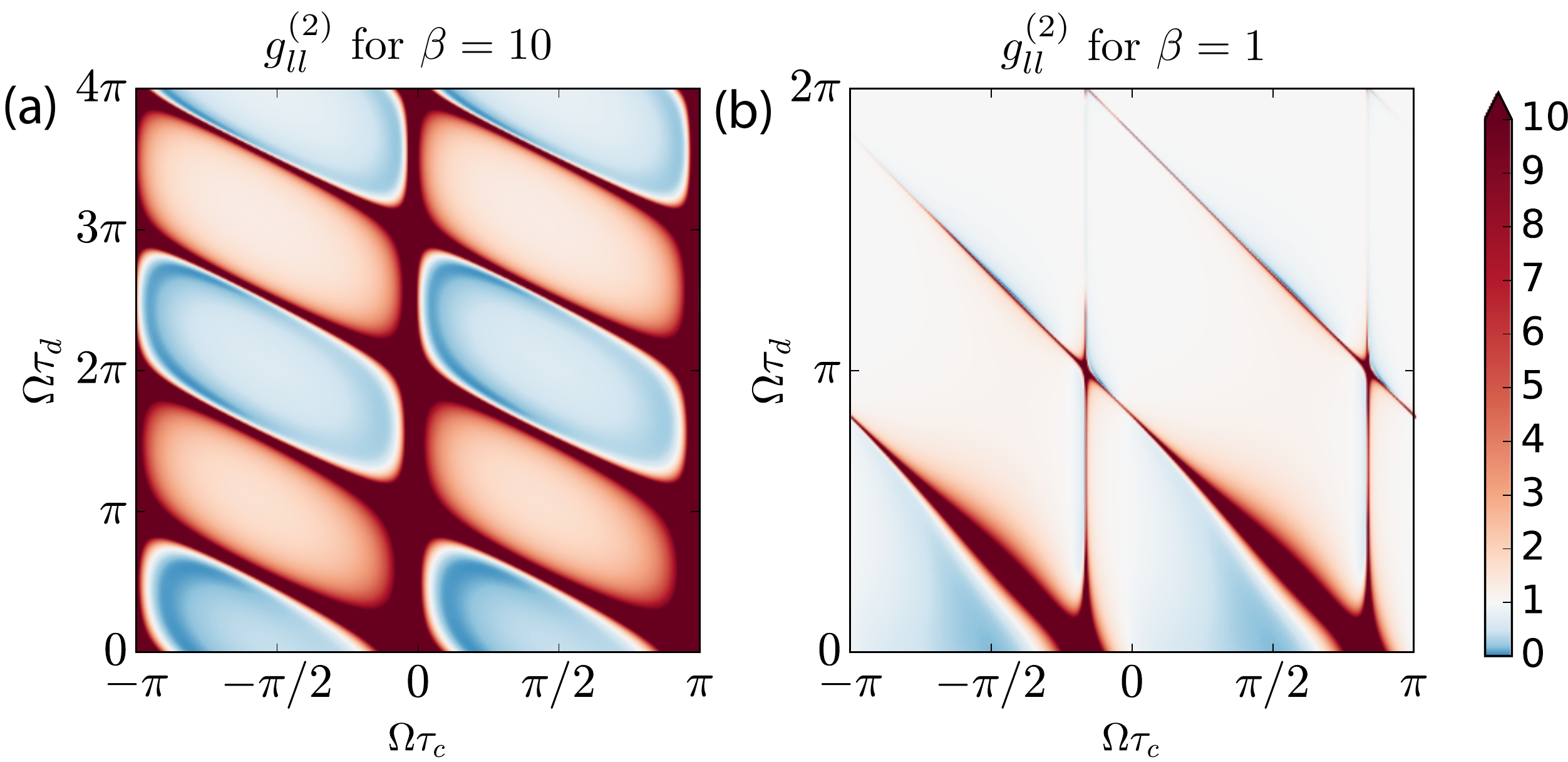}
	\caption{(Color online) Second order coherence of the reflected pulse $g_{ll}^{(2)}$ in  the ``sign change'' protocol as a function of central time $\tau_c$ and delay time $\tau_d$ for the Kerr  nonlinearity $|U|=4 \Gamma^{(0)}$. We show results for \textbf{(a)} the fast drive $\beta = 10$, where huge periodically repeated bunching peaks are formed and interwoven with areas of moderate bunching and anti-bunching; \textbf{(b)} the intermediate drive $\beta = 1$, showing the decay of $g_{ll}^{(2)}$ to the uncorrelated value (white area).}
	\label{fig:g2}
\end{figure}

Only fast drives, $\beta = \Omega/\Gamma^{(0)} \gg 1$, are able to affect the correlations before they decay, and we numerically calculate $g_{ll}^{(2)}$ for fast and moderate drives in the two protocols.
In the ``on-off'' protocol, the fast drive only induces small oscillations in the correlation function around the non-driven results \cite{seeSupple}. In contrast, the ``sign change'' protocol induces huge bunching effects due to the additional node in the single-photon reflection, as can be clearly seen in Fig.~\ref{fig:g2}(a). We also find periodic oscillations between strong bunching (red areas) and anti-bunching (blue areas) away from $\Omega \tau_c = 0$ and $\Omega \tau_c = \pm \pi$. This is a dramatic change in statistical properties of the reflected light due to the time dependence of $g (t)$ as compared to the case of constant $g$, where $g_{ll}^{(2)}$ is monotonously anti-bunched. For a moderate drive, $\beta = 1$, all oscillatory effects in the ``sign change'' protocol die out for delay times longer than a single drive period, as shown in Fig.~\ref{fig:g2}(b).

\emph{Summary.---}
We have proposed the quantum analogue of an optical chopper, operating at a few-photon level and realizable by a time-periodic modulation of the light-matter coupling strength. We have developed an exact Floquet scattering approach based on diagrammatic scattering theory and applied it to quantitatively describe scattering of microwave photons from the transmon qubit in two driving protocols of the coupling: ``on-off'' and  ``sign change''. In both of them we have observed interesting non-adiabatic memory effects arising due to the  driving. In particular, the ``on-off'' protocol produces periodic compressions of the photon's wavepacket, while the ``sign change'' protocol gives rise to the additional nodes in the envelope at which the field is completely quenched. Thus, in both cases the optical chopping is realized at the quantum single-photon level. In addition, in the ``sign change'' protocol we find dramatic changes in statistical properties of the reflected field showing up as strong bunching peaks in the $g^{(2)}$ function that are interwoven with periodically alternating areas of antibunching and moderate bunching --- features that are in sharp contrast to their non-driven counterparts. Our findings can be used in few-photon pulse manipulation, the design of photonic statistics at will and, as a consequence, in quantum information processing.

\emph{Acknowledgements.---}We are grateful to A. Fedorov and M. Hafezi for 
useful discussions. The work of V. G. is a part of the Delta-ITP consortium, 
a program of the Netherlands Organization for Scientific Research
(NWO) that is funded by the Dutch Ministry of Education,
Culture and Science (OCW).

\pagebreak
\widetext

\begin{center}
\textbf{\large Quantum choppers \\ Supplemental Material}
\end{center}

\setcounter{equation}{0}
\setcounter{figure}{0}
\setcounter{table}{0}
\setcounter{page}{1}
\makeatletter
\renewcommand{\theequation}{S\arabic{equation}}
\renewcommand{\thefigure}{S\arabic{figure}}
\renewcommand{\bibnumfmt}[1]{[S#1]}
\renewcommand{\citenumfont}[1]{S#1}

\section{Details of the model}

Following Ref.~[26] we provide an explicit form of the Josephson ring modulator Hamiltonian in terms of the modes $\Phi_X$, $\Phi_Y$, and $\Phi_Z$
\begin{eqnarray}
H_{JRM} &=& - E'_J \sum_{i=a,b,c,d} \cos \left(\delta_i + \frac{\pi \Phi}{2 \Phi_0} \right) + \frac{(\Phi_5 - \Phi_1)^2 + (\Phi_5 - \Phi_2)^2}{L_a} +  \frac{(\Phi_6 - \Phi_3)^2 + (\Phi_6 - \Phi_4)^2}{L_b} + I_p (\Phi_5 - \Phi_6) \nonumber \\
&=& - 4 E'_J \left( \cos \frac{\pi \Phi}{2 \Phi_0} \cos \frac{\pi \Phi_X}{\Phi_0}\cos \frac{\pi \Phi_Y}{\Phi_0} \cos \frac{\pi \Phi_Z}{\Phi_0}
+ \sin \frac{\pi \Phi}{2 \Phi_0} \sin \frac{\pi \Phi_X}{\Phi_0} \sin \frac{\pi \Phi_Y}{\Phi_0} \sin \frac{\pi \Phi_Z}{\Phi_0}\right) \nonumber \\
&+& \frac{2 \Phi_l^2}{L_a} + \frac{\Phi_X^2}{2 L_a} + \frac{2 \Phi_r^2}{L_b} + \frac{\Phi_Y^2}{2 L_b} +  I_p \left(\Phi_l - \Phi_r - \frac{\Phi_Z}{2} \right).
\end{eqnarray}

Setting $\Phi = \Phi_0 /2$ and expanding the nonlinear term in $\Phi_X/\Phi_0 , \Phi_Y /\Phi_0 , \Phi_Z /\Phi_0 \ll 1$, we obtain Eq. (1) [omitting the terms containing the decoupled modes $\Phi_l$ and $\Phi_r$].

\section{Details of the Floquet scattering formalism}

Let us consider an arbitrary Hamiltonian
\begin{equation}
H = H_0 + V (t),
\end{equation}
where the interaction $V (t)$ has a periodic time dependence $V (t) = V (t+T)$, and therefore can be expressed by the Fourier series
\begin{equation}
V (t) = \sum_{m} V^{(m)} e^{ - i m \Omega t}, \quad \Omega = \frac{2 \pi}{T}.
\end{equation}
A particular case is given by $V (t) = g (t) v_0$, where $g (t)$ is a time-dependent coupling strength.

To obtain a scattering operator we start from an equation for the evolution operator in the interaction picture
\begin{equation}
i\frac{d U_{\mathrm{int}} (t, t_0)}{d t} = V_{\mathrm{int}} (t) U_{\mathrm{int}} (t,t_0),
\label{diff_eq}
\end{equation}
where $V_{\mathrm{int}} (t) = e^{i H_0 t} V (t) e^{-i H_0 t}$. Taking the limit $t_0 \to -\infty$ we transform \eqref{diff_eq} into the integral form
\begin{equation}
 U_{\mathrm{int}}  (t) = \hat{1} - i \int_{-\infty}^t d t' e^{\eta t'} V_{\mathrm{int}} (t') U_{\mathrm{int}} (t'),
\label{int_eq1}
\end{equation}
or
\begin{equation}
U_{p' p} (t) \equiv \langle p' |  U_{\mathrm{int}}  (t) | p \rangle =  \delta_{p' p} - i \int_{-\infty}^t d t ' \sum_q \sum_m e^{ i (\varepsilon_{p'} - \varepsilon_q - m \Omega - i \eta) t'} V^{(m)}_{p'q} U_{q p} (t'),
\label{int_eq2}
\end{equation}
where an infinitesimal factor $\eta>0$ is additionally introduced for convergence.

We are looking for a solution of \eqref{int_eq2} at times $t>0$, satisfying the condition $\eta t \ll 1$, in the form
\begin{equation}
U_{p' p} (t) = \delta_{p'p} - \sum_{m'} \frac{e^{i (\varepsilon_{p'} - \varepsilon_p - m' \Omega) t}}{\varepsilon_{p'} - \varepsilon_p - m' \Omega - i \eta} \Theta_{p' p}^{(m')},
\label{ansatz}
\end{equation}
where $\Theta_{p' p}^{(m')}$ are constant matrices. Plugging \eqref{ansatz} into \eqref{int_eq2}, we obtain the equation 
\begin{equation}
\Theta_{p' p}^{(m')}= V_{p' p}^{(m')} -  \sum_q \sum_{n}  \frac{V^{(m'-n)}_{p'q} \Theta_{q p}^{(n)}}{\varepsilon_q - \varepsilon_p - n \Omega - i \eta} ,
\end{equation}
from which we can establish the matrices $\Theta^{(m')}$.

In the long time limit we make in \eqref{ansatz} the standard replacement $\frac{e^{i \omega t}}{\omega - i \eta} \to 2 \pi i \delta (\omega)$, and thus obtain the scattering matrix
\begin{equation}
S_{p' p} =  \delta_{p'p} - 2 \pi i \sum_{m'} \delta (\varepsilon_{p'} - m' \Omega - \varepsilon_p)  \Theta_{p' p}^{(m')}.
\label{Spp}
\end{equation}

Next, we introduce the matrix $T_{p'p}^{(m)} (E)$ obeying the equation
\begin{eqnarray}
T_{p' p}^{(m')} (E)= V_{p' p}^{(m')} +  \sum_q \sum_{n}  \frac{V^{(m'-n)}_{p'q} T_{q p}^{(n)} (E)}{  E - (\varepsilon_q - n \Omega) + i \eta} ,
\label{Teq1}
\end{eqnarray}
which is equivalent to Eq. (2).  $T_{p'p}^{(m)}$ coincides with the matrix $\Theta_{p' p}^{(m')}$ at the argument's value $E=\varepsilon_p$. Thus, we arrive at the expression
\begin{equation}
S_{p' p} =  \delta_{p'p} - 2 \pi i \sum_{m'} \delta (\varepsilon_{p'} -m' \Omega - \varepsilon_p)  T_{p' p}^{(m')} (\varepsilon_p) ,
\label{Spp1}
\end{equation}
which relates the $S$ matrix to the $T$ matrix in the time-periodic case. 

Applying to Eq. (2) the diagrammatic rules of Ref. [35] extended by additional summations over the Floquet indices, we obtain a closed expression for the normal ordered $N$-photon $T$ matrix
\begin{align}
&T^{m' m}_N (E) = \sum_{\{m'_j\}, \{m_j\}} P_0 \, \vdots \, V^{m' m^{\phantom{'}}_1} \tilde{G}^{m^{\phantom{'}}_1 m'_1} (E) V^{m'_1 m^{\phantom{'}}_2} \nonumber \\
&\times \ldots V^{m'_{2N-2}, m^{\phantom{'}}_{2N-1}}  \tilde{G}^{m^{\phantom{'}}_{2N-1},m'_{2N-1}} (E) V^{m'_{2N-1},m } \, \vdots \, P_0 ,
\label{TNormalmain}
\end{align}
given by the alternating product of $2N$ interaction operators, $V$, and $2 N-1$ dressed Green's functions, ${\tilde{G} (E) = (E-H'_0 - \Sigma)^{-1}}$, of the cavity. Here ${P_0 = |l=0 \rangle \langle l=0|}$ is a projector onto the dark state of the cavity, $H'_0 = H_0 - i \pd_{\tau}$ is the free Floquet Hamiltonian, and the Floquet components of the cavity's self-energy are found from $\Sigma^{mm'} \equiv \Sigma^{(m-m')} = - i \pi b^{\dagger} b \sum_n g^{(m-n)} g^{(n-m')}$. The symbol $\vdots (\ldots ) \vdots$ denotes the modified normal ordering, described in the main text.

To transform \eqref{TNormalmain} into Eq. (3), we express it first in the local time representation
\begin{eqnarray}
T_{N\varepsilon} (\tau) &\equiv & \sum_{m'} T_N^{(m')} (E) e^{-i m' \Omega \tau} =  \int_0^T \frac{d \tau_1}{T} \ldots \frac{d \tau_{2N}}{T} \delta_T (\tau -\tau_1) \nonumber \\
& \times &  P_0 \left( \vdots V (\tau_1) \tilde{G}_{\varepsilon} (\tau_1, \tau_2) V (\tau_2) \ldots V (\tau_{2N-1})  \tilde{G}_{\varepsilon} (\tau_{2N-1},\tau_{2N}) V (\tau_{2N}) \vdots \right) P_0 ,
\label{Ttau}
\end{eqnarray}
where we introduced the notations $\varepsilon=H_0-E = H_0 - \varepsilon_p$ and
\begin{equation}
\tilde{G}_{\varepsilon} (\tau , \tau') = \sum_{m,m'} e^{-i m \Omega \tau} \tilde{G}^{m m'} (E) 
e^{i m' \Omega \tau'},
\label{Gtt}
\end{equation}
and used the Poisson resummation formula
\begin{equation}
\sum_{m'} e^{-i m' \Omega (\tau-\tau_1)} = T \sum_{n} \delta (\tau - \tau_1 - n T) \equiv \delta_T (\tau - \tau_1).
\end{equation}
Then, from \eqref{Spp1} and \eqref{Ttau} we deduce that the $N$-photon scattering operator equals
\begin{equation}
S_N = \hat{1}_N  - i \int_{-\infty}^{\infty} d t e^{i (\varepsilon_{p'} - \varepsilon_p) t} T_{N \varepsilon} (t).
\end{equation}

It only remains to establish explicitly $\tilde{G}_{\varepsilon} (\tau , \tau')$ defined in \eqref{Gtt}. From the relations defining $G^{mm'} (E)$
\begin{eqnarray}
\sum_{m''} \left[ (m \Omega - \varepsilon) \delta_{m m''} - \Sigma^{m m''} \right] \tilde{G}^{m'' m'} (E) &=& \delta_{mm'}, \\
\sum_{m''} \tilde{G}^{m m''} (E) \left[ (m' \Omega - \varepsilon) \delta_{m'' m'} - \Sigma^{m'' m'} \right]  &=& \delta_{mm'},
\end{eqnarray}
we obtain the differential equations
\begin{eqnarray}
(i \pd_{\tau} - \varepsilon) \tilde{G}_{\varepsilon} (\tau, \tau')- \Sigma (\tau) \tilde{G}_{\varepsilon} (\tau, \tau') &=& \delta_T (\tau - \tau'), \\
(-i \pd_{\tau'} - \varepsilon) \tilde{G}_{\varepsilon} (\tau, \tau')-  \tilde{G}_{\varepsilon} (\tau, \tau') \Sigma (\tau') &=& \delta_T (\tau - \tau'),
\end{eqnarray}
where $\Sigma (\tau) = - i \pi g^2 (\tau) b^{\dagger} b$. Being equipped with the periodic boundary conditions in both variables, they acquire the following solution
\begin{equation}
\tilde{G}_{\varepsilon} (\tau, \tau') = - i T \sum_n \Theta (\tau - \tau' - n T) e^{-i \bar{\varepsilon} (\tau - \tau' - n T)} e^{-F_{\mathrm{osc}} (\tau) + F_{\mathrm{osc}} (\tau')},
\end{equation}
where $\bar{\varepsilon}  = \varepsilon + \Sigma^{(0)}$ and $F_{\mathrm{osc}} (t) = - \sum_{m \neq 0} \frac{\Sigma^{(m)}}{m \Omega} e^{- i m \Omega t}$.
Inserting it into \eqref{Ttau} and extending the finite integration ranges $0 < \tau_j <T$ to the infinite ones $-\infty < t_j < \infty$, we obtain Eq. (3).

\section{The intensity-intensity correlation function}

\begin{figure}[ht]
\begin{minipage}[t]{0.45\linewidth}
\centering
\includegraphics[width=\textwidth]{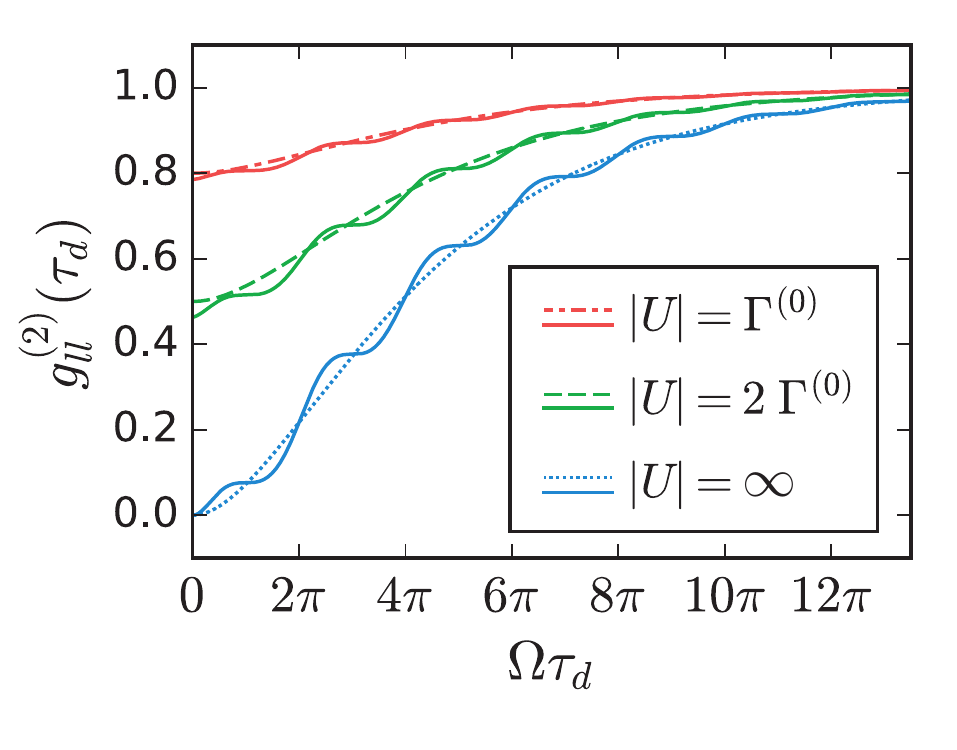}
\caption{The $g^{(2)}_{ll}(\tau_c = 0, \tau_d)$ correlation function for the ``on-off'' protocol at fast driving, $\beta=10$. The $g^{(2)}$ correlation for the corresponding non-driven system with a decay rate set to $\Gamma^{(0)}$ are shown as dashed lines, and the (uninteresting) correlations for the driven system slightly oscillate around the non-driven antibunching curves.}
\label{fig:g21}
\end{minipage}
\hspace{0.2cm}
\begin{minipage}[t]{0.45\linewidth}
\centering
\includegraphics[width=\textwidth]{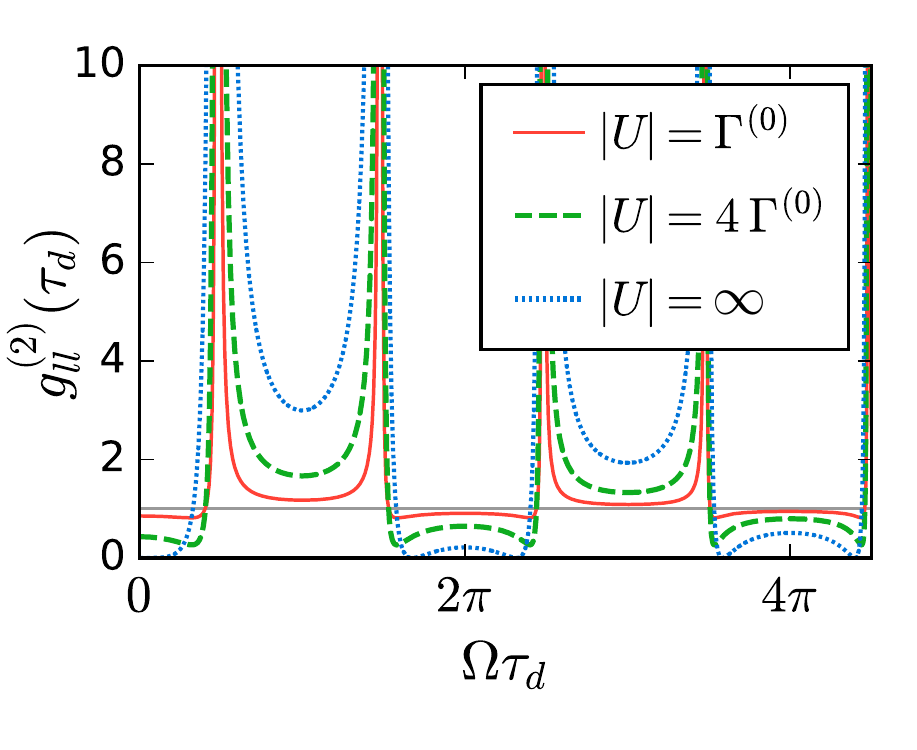}
\caption{The $g^{(2)}_{ll}(\Omega \tau_c = -\pi/2, \tau_d)$ correlation function for the ``sign change'' protocol at fast driving, $\beta=10$, for various Kerr nonlinearities $U$. The alternating bunching and anti-bunching is evident, with the correlations increasing with increasing $U$.}
\label{fig:g22}
\end{minipage}
\end{figure}

For completeness we show additional numerical results for the $g^{(2)}$ correlation function. The ``on-off'' case only induces small oscillations on top of the non-driven correlations, as can be seen from the $g^{(2)}_{ll}(\tau_c = 0, \tau_d)$ plot shown in Fig.~\ref{fig:g21}. For comparison we also plot the correlation function for the non-driven system fixing the decay rate to the time-averaged value of the driving protocol.

For clarity Fig.~\ref{fig:g22} showcase $g^{(2)}_{ll}(\Omega \tau_c = -\pi/2, \tau_d)$ for the ``sign change'' protocol at fast driving for various values of the Kerr nonlinearity, $U$, hence presenting a line-cut of Fig.~3(a) from the main text.

\begin{figure}[tb]
	\centering
	\includegraphics[width=.5\textwidth]{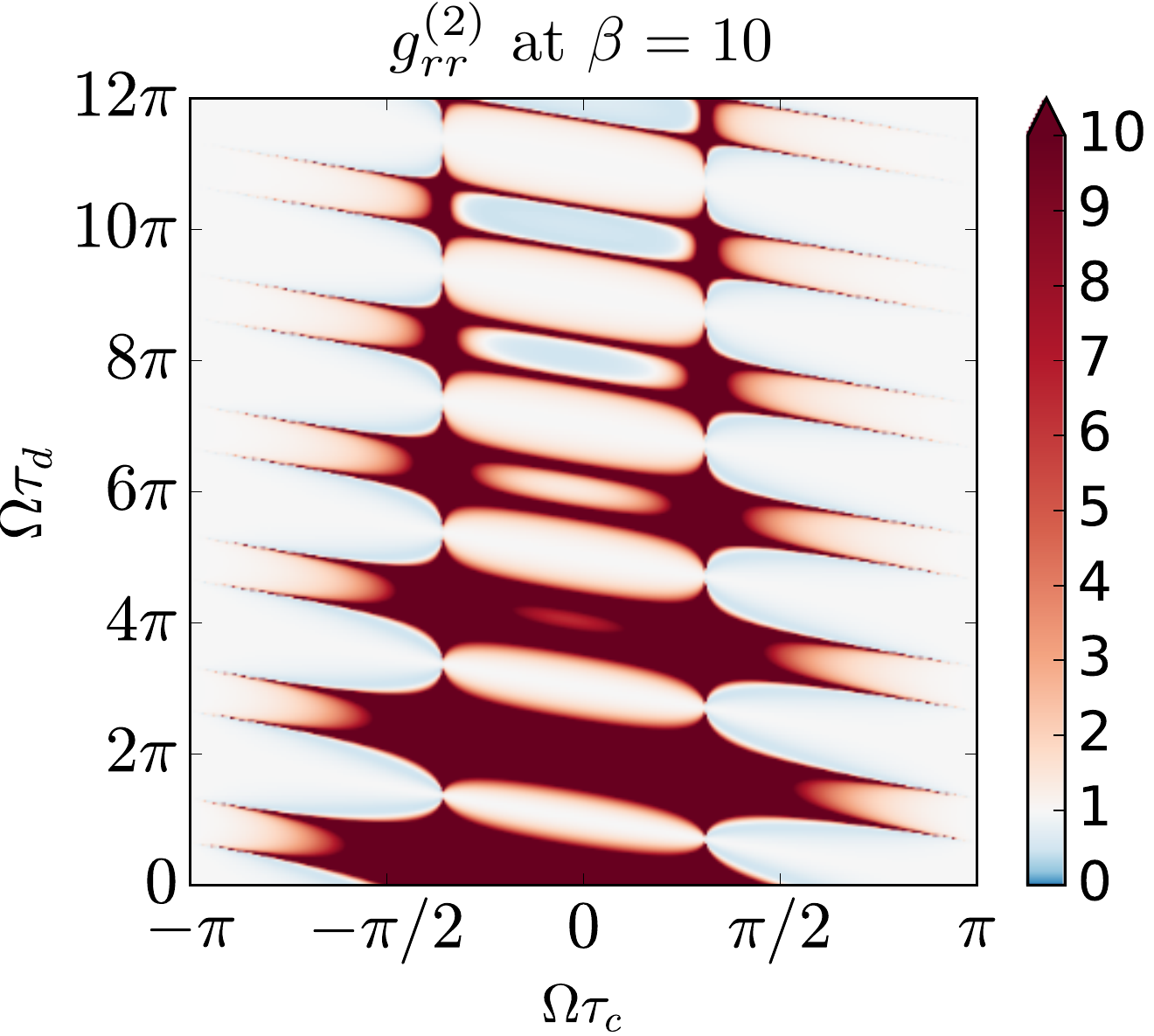}
	\caption{The $g^{(2)}_{rr}$ correlation of transmitted light in the ``on-off'' protocol at fast driving, $\beta = 10$, with a nonlinearity $|U|=2\Gamma^{(0)}$. Note the strong periodically recurring bunching due to the wavepacket compression.}
	\label{fig:g23}
\end{figure}

The photon compression by the ``on-off'' driving introduces nodes in the transmission and produces, similarly to the field quench effects in the reflected light for the ``sign change'' protocol, strong bunching in the transmitted light captured by $g^{(2)}_{rr}$. This picture is verified by a numerical calculation of the correlation function for fast drive, $\beta=10$, and nonlinearity, $|U| = 2 \Gamma^{(0)}$ as shown in Fig.~\ref{fig:g23}.

\end{document}